\documentclass[conference]{IEEEtran}
\usepackage{booktabs} % nicely formatted tables
\usepackage{url}
\usepackage{soul,color,graphicx,float}
\usepackage{tablefootnote}
\usepackage{eurosym,booktabs,multirow}
\usepackage{subcaption}
\usepackage{mathtools}
\usepackage{graphicx}
\definecolor{Orange}{rgb}{1,0.5,0}

\begin{document}
\title{Standing on the Shoulders of Giants: AI-driven Calibration of Localisation Technologies}

\author{\IEEEauthorblockN{Aftab Khan\IEEEauthorrefmark{1}\IEEEauthorrefmark{2},
Tim Farnham\IEEEauthorrefmark{2},
Roget Kou\IEEEauthorrefmark{2}\IEEEauthorrefmark{3},
Usman Raza\IEEEauthorrefmark{2},
Thajanee Premalal\IEEEauthorrefmark{4},\\
Aleksandar Stanoev\IEEEauthorrefmark{2},
William Thompson\IEEEauthorrefmark{2}}\\ 
\IEEEauthorblockA{\IEEEauthorrefmark{2}Toshiba Research Europe Ltd., Bristol Research \& Innovation Laboratory, Bristol, UK\\ 
\IEEEauthorrefmark{3}University of Bristol, Bristol, UK\\
\IEEEauthorrefmark{4}University of Leeds, Leeds, UK\\
Email: aftab.khan@toshiba-trel.com}}
\maketitle

\begin{abstract}
High accuracy localisation technologies exist but are prohibitively expensive to deploy for large indoor spaces such as warehouses, factories, and supermarkets to track assets and people. However, these technologies can be used to lend their highly accurate localisation capabilities to low-cost, commodity, and less-accurate technologies. In this paper, we bridge this link by proposing a technology-agnostic calibration framework based on artificial intelligence to assist such low-cost technologies through highly accurate localisation systems. A single-layer neural network is used to calibrate less accurate technology using more accurate one such as BLE using UWB and UWB using a professional motion tracking system. On a real indoor testbed, we demonstrate an increase in accuracy of approximately 70\% for BLE and 50\% for UWB. Not only the proposed approach requires a very short measurement campaign, the low complexity of single-layer neural network makes it ideal for deployment on constrained  devices typically for localisation purposes.
\end{abstract}

\section{Introduction}

%Recent advances in hardware miniaturisation..........,
%indoor localisation 
%have made indoor localisation more practical 
%and 
Indoor localisation market is expected to experience steady growth and is forecasted to hit \$29.4 billion in revenue by 2022 from its \$3.43 billion shares in 2015 ~\cite{reuters}. A number of wireless localisation technologies do exist, some more accurate than others due to their design choices such as carrier bandwidth, number of antennas, number of radio samples, etc. To offer an example, it is hard for a narrow-band technology like Bluetooth Low Energy (BLE), which is popular with handset devices and thus is becoming cheaper day-by-day due to economy of scale, to match up the accuracy of Ultra-Wideband (UWB) technologies like Decawave that make use of up to 500 times more bandwidth and is less vulnerable to multi-path fading effects. However, the fact that regulations on UWB technologies are stricter across different regions (such as Japan, America, and Europe) there is a very good case for resorting to narrowband technologies.  

%Furthermore, UWB technologies are subject to more strict regional regulations than BLE that operates in the global 2.4GHz Industrial, Scientific, and Medical (ISM) band.

As the accuracy comes at the expense of higher cost and energy consumption, it would be ideal to use pocket- and battery-friendly commodity technologies and yet achieve or approach the accuracy of relatively more expensive and power-hungry technologies. In this paper, we try to achieve exactly this objective by leveraging latter to \emph{calibrate} the former in a similar way in which low-cost and cheap sensors are calibrated using more specialised, expensive industrial grade sensing instrumentation. 
We propose a novel machine learning based approach referred to as ICON\footnote{ Intelligent CalibratiON} that trains neural networks using measurements as features from low-cost technology and location estimates from the accurate technology based on data collected from a deployment area. Different from legacy finger-printing solutions, ICON does not require a huge amount of data to improve localisation accuracy. Approximate location and small training campaigns work very well. Once the training phase is completed, the generated model can be deployed to improve localisation accuracy on low-cost technology without the need for expensive technology.  

In this paper, we not only propose ICON -- our technology-agnostic calibration method -- but also build a real-world indoor testbed that employs two key radio technologies including BLE and Decawave UWB and a highly accurate commercial motion capture technology from OptiTrack. As these achieve a very different level of accuracy, it enables us to calibrate one against the other. Our detailed experiments show that calibration through ICON achieves a statistically significant increase in localisation accuracy. ICON approximately halves the localisation error of UWB when calibrated against motion capture technology, while it reduces it by approximately 70\% for BLE when calibrated against UWB. These gains can be achieved solely by incorporating the intelligent neural network model in the software running on localisation technology. No additional hardware-based optimisations such as the use of multiple antennas or accurate crystal clocks are required.

\begin{figure*}[tp]
    \centering
    \includegraphics[width=0.8\textwidth]{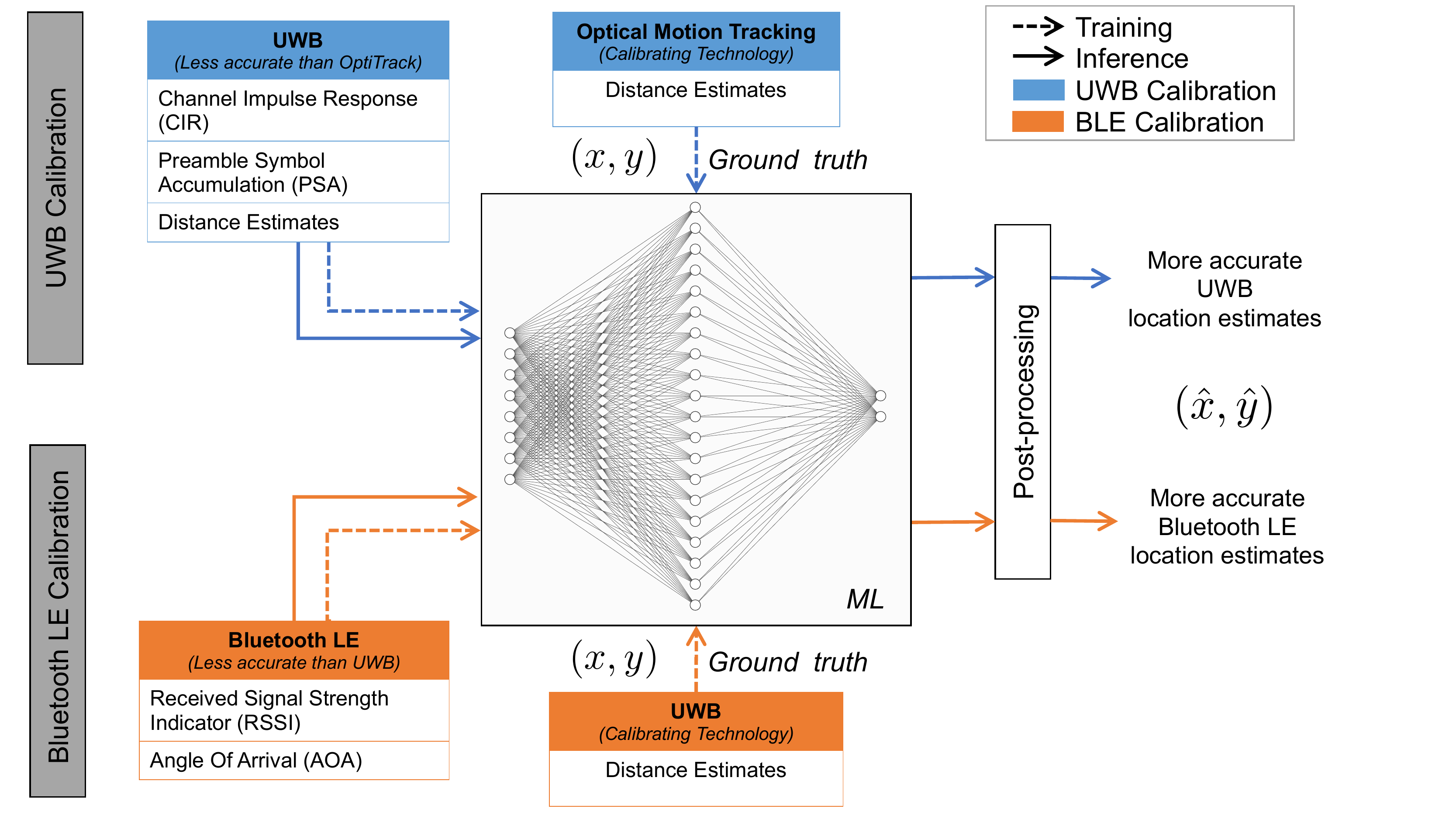}
    \caption{Overview of the proposed intelligent calibration method separately illustrating UWB and BLE calibration in blue and orange respectively.}
    \vspace{-0.5cm}
    \label{fig:overview}
\end{figure*}

\section{Background}

Real time location and tracking systems can be supported using many different technologies. This includes a range of radio and image based approaches that have different trade-offs in terms of cost, accuracy and power consumption. For asset tracking use-cases, the tags must be low-cost and must have a long battery life, so it is attractive to use technologies that permit passive tags or tags with very low power consumption. This includes RFID \cite{526899}, Bluetooth Low Energy (BLE) \cite{whitepaper}, WiFi \cite{Kotaru:2015:SDL:2785956.2787487} and image based tracking approaches.

One of the most attractive radio technologies for low cost asset tracking is BLE due to the low power consumption at a reasonable range. The Angle of Arrival (AoA) estimation capabilities, that have recently been standardised in Bluetooth 5.1 \cite{whitepaper}, can enhance the ability to accurately track assets. The scenarios that are particularly challenging for this technology are within complex multipath radio propagation environments, such as in offices, retail or industrial deployments.

An alternative technology is Ultra-WideBand (UWB), which mitigates the frequency selective fast fading by using a large bandwidth transmission. This is typically around 500MHz, which permits Time of Arrival (ToA) or Time Difference of Arrival (TDoA) techniques to be applied in order to estimate the distances from targets to locators. However, as this technique requires accurate timing information, it normally relies on a handshake between each locator and corresponding target. 

Passive RFID localisation \cite{526899} can also exploit AoA and signal strength based approaches in a similar way to BLE. However, as passive RFID relies on the ability to detect individual backscatter signals from each tag there is a range and density related limitation due to weak return signals and interference (or collision) between them.  Therefore, achieving a reasonable range and localisation accuracy is still a major unresolved challenge.

Finally, many different image based tracking techniques are now often used for asset tracking. This can utilise optical or depth image cameras that can resolve the distance to the targets using Time of Flight (ToF) or stereo camera techniques. These approaches (such as using the infrared Kinect V2 ToF camera) can provide a good depth resolution at reasonable cost. However, the tracking of targets can only be reliably performed over limited distances and with unobscured line of sight. Therefore, combining of radio and image based techniques is an attractive option to overcome these limitations.

\section{ICON -- ML Driven Accurate Calibration}
\label{sec:method}

Overview of the proposed framework is illustrated in Figure \ref{fig:overview}. In this, we show the ML-driven calibration of the two scenarios (BLE calibration through UWB and UWB calibration through a motion capture system). 
In the case of BLE calibration (highlighted in orange), we used RSSI and AoA measurements as features for training the ML model. For UWB calibration, we used channel impulse response (CIR), preamble symbol accumulation (PSA) and distance estimates. 
A generalisable test bed was created in order to test the proposed system detailed in Section \ref{sec:testbed}.

In this paper, we used a supervised regression modelling approach (using Neural Networks in particular). 
In theory, other regression approaches can also be used for this purpose, however Neural Networks are known to perform well in such scenarios e.g., AoA estimation \cite{8554304}; both in computational time (during inference/deployment), and accuracy.

\begin{figure*}[tp]
    \centering
    \includegraphics[width=0.75\textwidth]{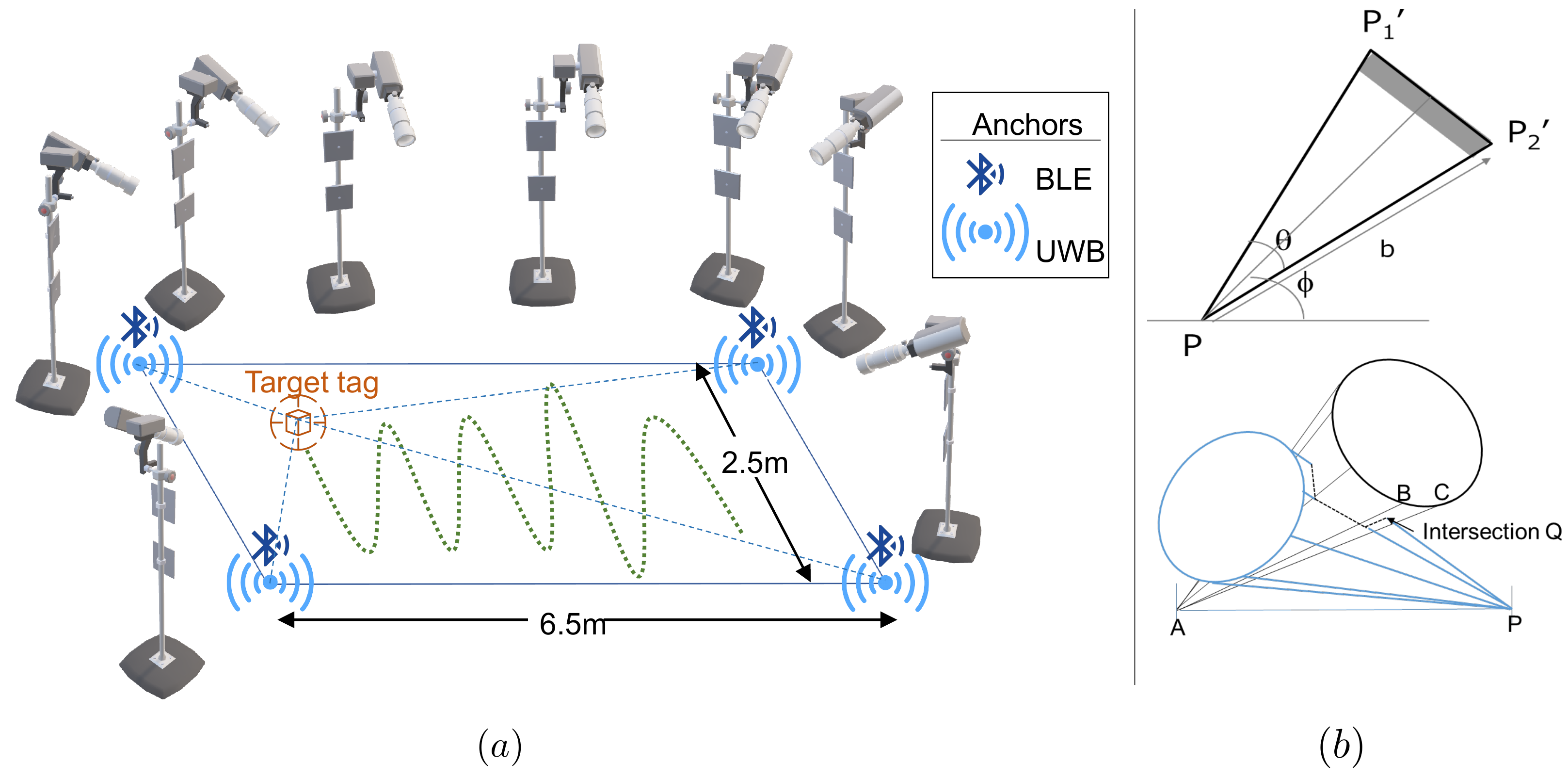}
    \caption{a) Illustration of the experimental testbed; b) AoA localisation  with AoA $\phi$ and $\theta$, and range estimate $b$ (top); with intersection between  AoA estimates (bottom).}
    \label{fig:testbed}
    \vspace{-0.5cm}
\end{figure*}

We used regression modelling in particular due to the nature of the problem since we are interested in the location estimates (that are essentially numerical) of the target tags.
Our feature vector, based on the measurements of both BLE and UWB can be formally defined as:

\begin{equation}
    \vec{f}_{\text{BLE}}  = (\text{RSSI},\text{AoA})^{d=16}
\end{equation}

\begin{equation}
    \vec{f}_{\text{UWB}}  = (\text{CIR},\text{PSA},\text{distances})^{d=12}
\end{equation}

For Neural Networks (NN) training, we used a single layer architecture which is computationally more efficient compared against other deep learning approaches; this is critical as deployment of these models may eventually be required on low-energy devices with limited computational capacity. For training, which is an off-line process (highlighted using dashed lines in Figure \ref{fig:overview}), we used Bayesian regularisation \cite{614194} with the number of input nodes depending on the two aforementioned scenarios (equal to $d$). We used $50$ hidden nodes, and $2$ output nodes representing a 2-D Cartesian location estimate.

For evaluation, we used a leave-one-session-out cross-validation scheme; the most preferred in such a context that aims to show the generalisation performance of the proposed framework. Experimental details related to this are explained in Section \ref{sec:exps}. The two dimensional output from the Neural Network is then post-processed to produce the resulting estimates  $(\hat{x},\hat{y})$. During post-processing, smoothing is performed using a moving average method (consistently used for both the baseline and predicted estimates). 
Baseline methodologies for both BLE and UWB calibration are introduced in the next section.
The estimated locations are then compared against the ground truth (UWB estimate for BLE calibration or motion capture location estimate for UWB calibration) using pairwise distances between the two. In particular, we used Euclidean distance to compute the estimation errors:

\begin{equation}
    e = \sqrt{(\hat{x}-x)^2+(\hat{y}-y)^2)}
\end{equation}

\section{Localisation Testbed}
\label{sec:testbed}

Two testbeds were created for data collection in this paper. First was used to collect data for Bluetooth calibration using Ultra-Wideband (UWB) and the second was designed in order to collect data for UWB calibration using a much superior motion capture  system. The second scenario required a different physical setup compared to the first setup. Combined illustration for the two scenarios is shown in Figure \ref{fig:testbed}.

%The 2 testbeds used to obtain training data as follows:
%\begin{itemize}
%  \item Bluetooth with Ultra-Wideband
%  \item Ultra-Wideband with Optical motion tracking system
%\end{itemize}

\subsection{Bluetooth Calibration using Ultra-Wideband Testbed}

\subsubsection{Bluetooth locators}
For BLE, the locator nodes were developed using GNU Radio software and Ettus Research Universal Software Radio Peripherals (USRP) N210 hardware. Since the aim of this work is to investigate the localisation performance of future AoA enabled BLE systems, the implemented system was developed to closely emulate the functionality of future BLE systems. 

The AoA functionality in the BLE direction estimation standard \cite{whitepaper} is enabled by the inclusion of an additional transmission frame element, called the constant tone extension (CTE). The CTE consists of a single frequency tone transmitted after the Cyclic Redundancy Check (CRC), with a duration between 16 and 160 $\mu s$. This is equivalent to between 16 and 160 symbols, when assuming a symbol rate of 1MHz. For these investigations a slightly modified frame is used. A TI CC2650 Launchpad development board was configured to transmit a conventional BLE packet, with a payload content all set to the same binary value, thus producing a constant tone payload, which shall act as the CTE. The rest of the packet was configured to be a non-connectable beacon device, with a transmission repetition of 50ms. Due to receiver hardware limitations, no frequency hopping was used, and the transmission centre frequency was set to 2.402GHz, i.e. BLE channel 37. The main differences between this and a final AoA BLE setup, is that here the CTE is before the CRC not after it, and no frequency hopping takes place.  

The locator nodes each use a four antenna element circular array, with elements spaced at 0.45$\lambda$, where $\lambda$ is the wavelength of the centre of the BLE transmission band, i.e. 2.44GHz. In AoA enabled BLE products, it is assumed that the devices will only contain a single RF chain, and during reception of the CTE the feed into the RF chain will be switched between the antenna array elements. Instead of this the switching is emulated using  four N210 USRPs, one for each antenna element, and the received data streams are then switched in software. 

During operation, the locator node captures any transmissions from the target node, extracting the payload, which acts as the CTE. To emulate the switched antenna system, the first eight payload symbols are retained from the first USRP, U1, these are used for frequency and timing synchronisation of the streams. The next 152 symbols are divided between the four USRPs in a round robin fashion. To emulate the finite switching time, after each symbol is allocated to one stream, the next symbol is disregarded. This results in each stream retaining one in 8 samples which are fed into the AoA algorithm, deployed on a PC.

The MUltiple SIgnal Classification (MUSIC) algorithm \cite{van2004optimum} is used to provide an angular spectrum, where the largest value represents the most likely AoA. When there are significant multipath reflections, there is a possibility for peaks either related to indirect paths or combination paths due to an inability to resolve them. During testing it was found that some AoAs  were significantly different from the neighbouring ones. The possible causes of these erroneous estimations are strong multi-path reflections from the environment, or reception of noisy CTE packets. A filter was used to average the past five AoA estimates, passing on the two estimates of the past five that are closest to the average to the localisation algorithm. 

\subsubsection{Bluetooth Localisation} \label{kalman}

In order to perform localisation of targets, the AoA estimated from multiple Bluetooth locators are triangulated. This permits one or more location candidates to be obtained based on the 3D intersections of the conical target AoA ($\phi$ in azimuth and elevation) estimated from the different locators. A 6 degree absolute path error $\theta$ is used, as shown in Figure \ref{fig:testbed}b (top), which is the expected AoA accuracy. However, as multi-path reflections result in multiple anomalous AoA paths the two most likely paths per locator are used. Many erroneous AoA path estimates are still obtained despite this initial filtering. Therefore, the intersection test attempts to further eliminate unfeasible candidates by representing the conical AoA path error regions as triangulated irregular networks (TINs), as shown in Figure \ref{fig:testbed}b (bottom). Barycentric coordinate computation is then used to obtain the boundary of the most likely candidate target location regions. The equations for the intersection tests are:

$s=  ((u.v)(w.v)-(v.v)(w.u))/(〖(u.v)〗^2-(u.u)(v.v))$

$t=  ((u.v)(w.u)-(u.u)(w.v))/(〖(u.v)〗^2-(u.u)(v.v))$

Intersection occurs if $ (s >= 0 \textrm{ and } t >= 0 \textrm{ and } s + t <= 1)$
where $u = A\rightarrow B$, $v = A\rightarrow C$ and $w = A\rightarrow Q$  and $Q$ is the intersection point on the plane $(A,B,C)$ with $A$, $B$ and $C$ being the clockwise ordered triangle face vertices representing the boundary of the AoA estimation region. 

The intersection test is performed for each of the two paths from pairs of locators with each of the vertices and TIN triangles forming the boundary of a 3D target location estimation region. Therefore, for two paths per locator there are four combinations per locator pair. Also, twelve vertices are used per path TIN region resulting in 48 tests per pair of locators. With four Bluetooth locators there are six locator pair combinations and so this results in 288 tests forming the boundaries of target location candidates. The expected target location is the centroid of the final candidate region after elimination of the unfeasible candidates. The computations can be efficiently performed on GPU hardware \cite{SKALA2008120} to permit central processing of AoA path intersections from all the locators.

After this, unfeasible candidates are further eliminated by firstly removing regions that are too far from the expected location based on the predicted location using a Kalman filter and then ranking the remaining candidate regions by the number of intersection overlaps observed between them. 

\subsubsection{Ultra-Wideband Locators}
For our ground truth against the BLE test setup, we used the Decawave's MDEK1001 development kit. 
This development kit includes a development board that has a Nordic Semiconductor nRF52832 SoC as well as Decawave's latest DWM1001 UWB module. 
The locators from Decawave have been designed with localisation features provided by Decawave's Positioning and Network Stack (PANS), which is a precompiled binary firmware for use with the MDEK1001 devices. 
The PANS software also includes Decawave's proprietary location engine which pairs with an Android application DRTLS which was used to setup a test environment for tracking a target; see Figure \ref{fig:testbed}.

\subsubsection{Data Collection}

The data collected for this testbed include:
\begin{itemize}
    \item Angle of Arrival data for Bluetooth
    \item Received Signal Strength Indicators %which give quantifiable values for the signal strength of the received signal from the beacons
    \item BLE Cartesian co-ordinates as detailed above
    \item Cartesian co-ordinates from UWB locators
\end{itemize}

%Our data collection layout was a $6.5$m by $2.5$m test area in which a single target was tracked. 

\subsection{Ultra-Wideband with OptiTrack Motion Capture Testbed}

\subsubsection{Ultra-Wideband Locators and Software}
As in the previous testbed, we used the Decawave MDEK1001 Development boards as our locators. 
However, to train an ML model, location data provided by Decawave's PANS software is not enough. 
To obtain raw measurement data, % about other features which can aid in localisation, 
we developed a custom firmware for the MDEK1001 boards.% in order to obtain the data we require. 
The custom firmware was built upon a port of the open-source RTOS Zephyr Project. 
This port, provided by RT-LOC, is a very comprehensive real-time operating system with low level functions and a precise clock that enables advanced implementation. 
Our software, is a single-sided two-way ranging application that is initiated by the target which sends out a poll message to a specified anchor. 
The target then waits for the anchor to send a response which contains the time stamps of the anchors, time stamps for when it received and sent the messages. 
This is thereby enough measurement data for the target to calculate time of flight of the messages and calculate a distance.

%In addition to distance the more features would be ideal in order to train the model. 
Unlike BLE, RSSI data is not readily available, therefore instead of obtaining the actual RSSI, the DWM1001 modules contain data relating to packet receive quality for debugging. 
We thereby modified the software to read from particular registers which can obtain the channel impulse response power (CIR) of the received signal and preamble symbol accumulation (PSA). The CIR is an indicator of the power of the received impulse from the burst that is a packet being received.%; this is similar to that of the received signal indicator. 
We also collected PSA, which is the number of preamble symbols accumulated on the receiver, as preamble symbols help receivers determine whether a signal is just noise or an actual message. 
The PSA allows us to know how many symbols were received and whether they were enough to allow the receiving device recognise that it was a message.

\begin{figure*}[tp]
	\centering
	\begin{subfigure}[t]{0.35\textwidth}
		\includegraphics[width=\columnwidth]{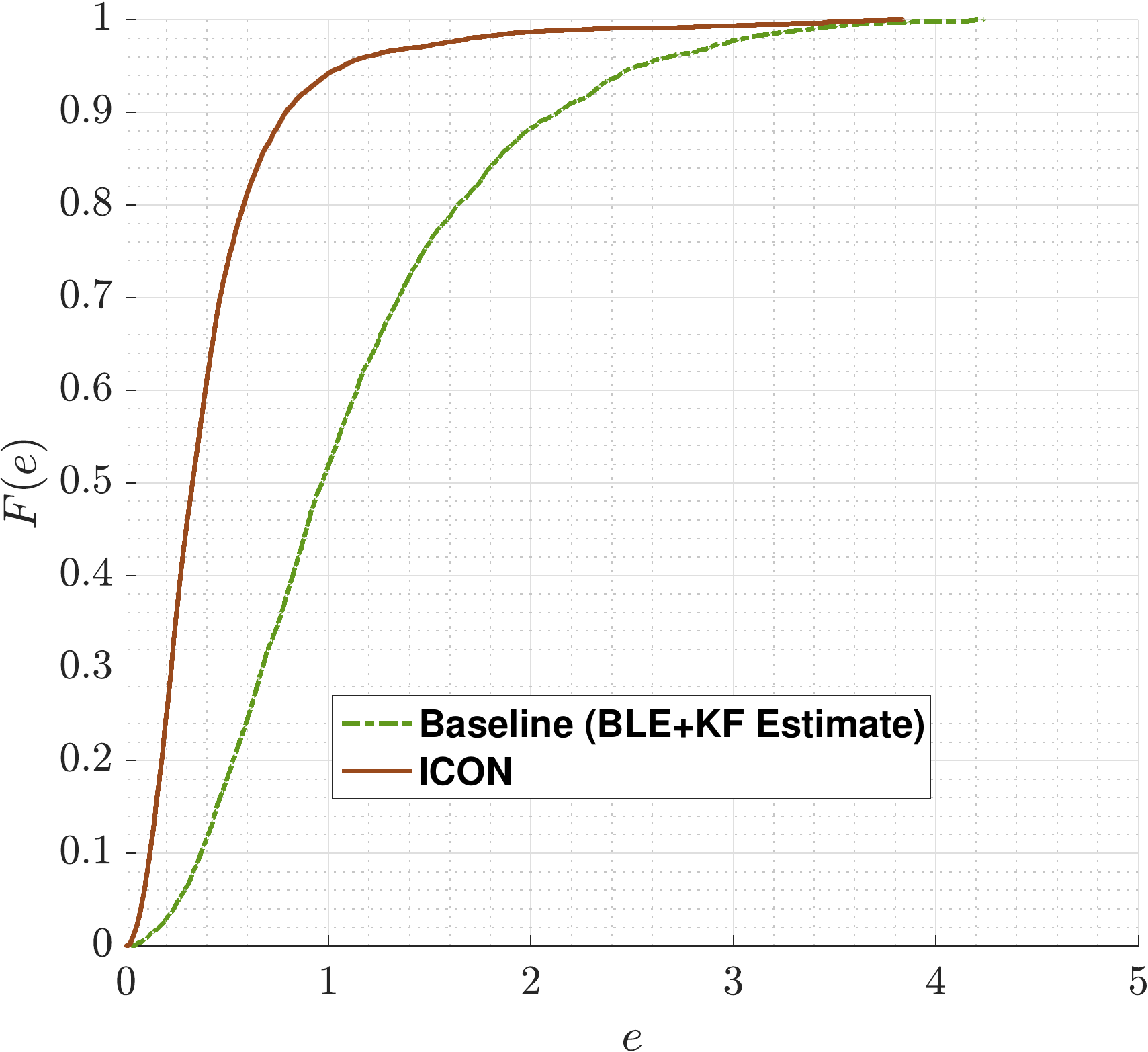}
		\caption{\label{fig:bleuwb}
		}
	\end{subfigure}
	\hspace{2cm}
	\begin{subfigure}[t]{0.35\textwidth}
		\includegraphics[width=\columnwidth]{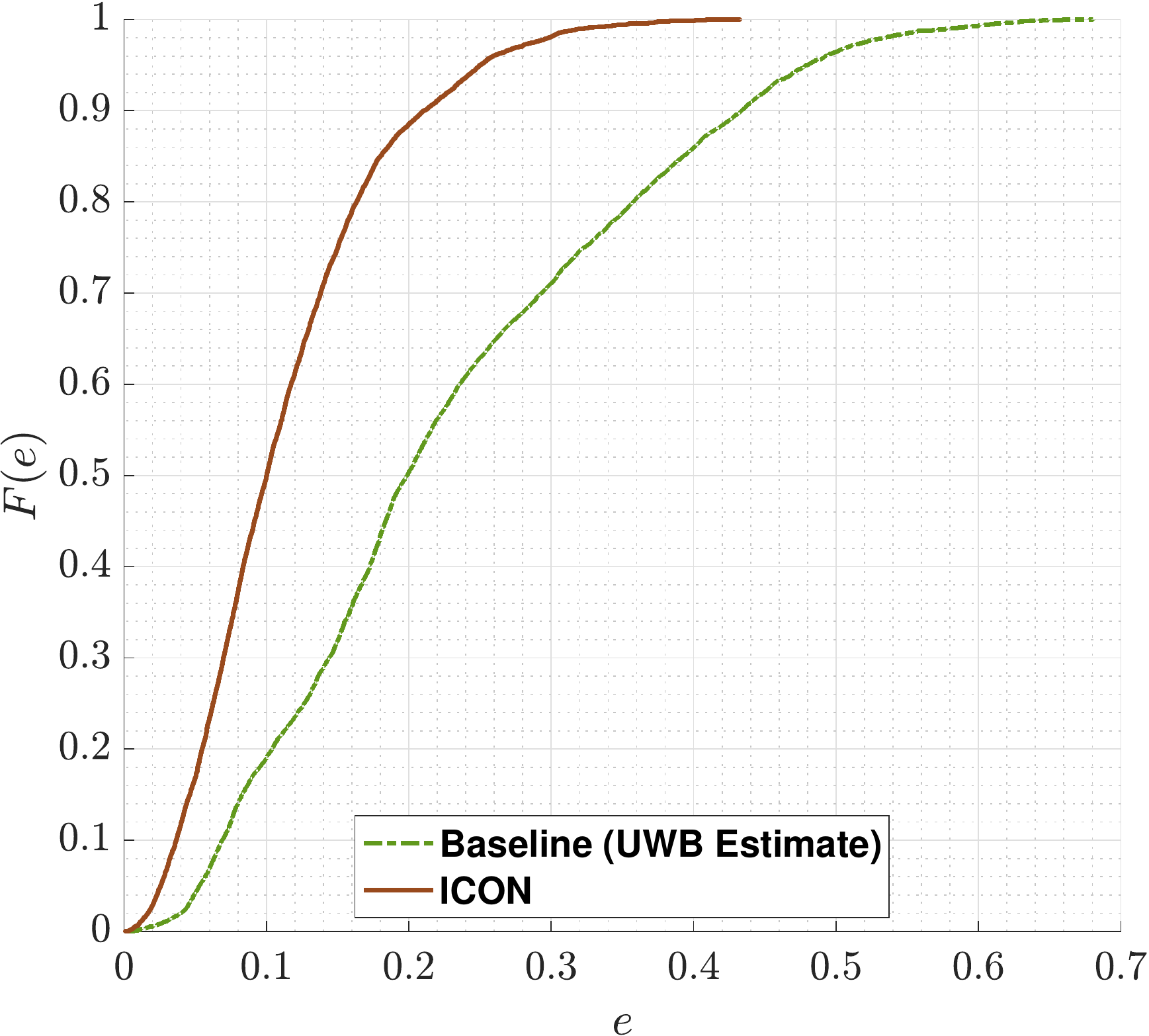}
		\caption{\label{fig:uwboptical}
		}
	\end{subfigure}
	\caption{\label{fig:results}
	(a) BLE calibration using UWB; (b) UWB calibration using Motion Tracking.
    }
    %\vspace{-0.5em}
\end{figure*}

However, receiving to and from one device is not particularly useful, therefore our testbed is using four anchors, the target has consequently a schedule to collect data from the four separate anchor devices on a time division basis. The target was configured to collect a set of data at a rate of $10$Hz in order to match that of the PANS localisation.

\subsubsection{Motion Capture System for Localisation}
In order to provide a ground truth against the UWB system, we used a motion capture system from OptiTrack\footnote{https://optitrack.com/}, using 8 cameras covering the target area (shown in Figure \ref{fig:testbed}). 
The 8 cameras track objects defined as \emph{rigid bodies}, which are small grey spheres attached to the target object. 
However, the motion capture system requires some prior calibration before use. 
Calibration is performed using the OptiTrack software and %require a process they call `wanding' in which there is a `wand' with 3 rigid bodies of set distance in which is tracked by the cameras to collect samples and develop an area map of the which they are set to track. So by moving the rigid bodies around within the capture zone of the cameras, the cameras collect samples of the different places the `wand' has been and calibrates the recording zone. 
with about 5-10 minutes of calibration, the software estimates localisation accuracy to have an error of 0.29mm which is significantly more accurate than the proclaimed decimetre level accuracy of UWB. 
Through the calibration process, we ensure that the origin points of the two systems exactly align.

\subsubsection{Data Collection}
It is very challenging to enable the PANS software to use additional functions in addition to the pre-built software. In order to overcome this, we used two anchors and two targets.
%One limitation of the PANS software was its i own functions on top of their software, so in order to work around this we used 2 anchors and 2 target tags. 
One set of four anchors and single tag was used for the location estimates using the DRTLS and PANS software provided by Decawave, whilst the other set of the identical setup was used for collecting the measurement data. 
%One problem of this would be that if the 2 setups were to communicate on the same channel then there would be packet collision causing errors, so to prevent we ensured the systems were on different UWB channels. 
In order to avoid packet corruption, we configured the two setups to communicate on different UWB channels.
When collecting the data, we connected the targets to a Raspberry Pi which logged both the Cartesian co-ordinates and the measurement data. 
%The Raspberry Pi also logged the time stamps and was clock synchronised over the internet. 
The motion capture location estimates were collected as Cartesian co-ordinates over time on a separate computer connected to the cameras with the OptiTrack software.
Clocks across all the three setups were synchronised over the internet.

The following measurement data was collected through this testbed:
\begin{itemize}
    \item Channel Impulse Response power indicator
    \item Preamble symbol accumulation on the target
    \item Distance to each anchor in metres
    \item Cartesian co-ordinates of the Ultra-Wideband target
    \item Cartesian co-ordinates of the motion capture target
    %\item Internet synchronised clock time stamps
\end{itemize}

\section{Experiments}
\label{sec:exps}

\subsection{Datasets} 
The data collection process through the aforementioned testbeds created 2 datasets. 
Both datasets covered two scenarios, \emph{i)} Walking: in this the target object was carried by a human subject, and \emph{ii)} Trolley: in this a trolley was used that carried the target object. In the later case, the trolley was remote controlled to move around the test area. Further details are provided below.

\subsubsection{Bluetooth/Ultra-Wideband}
The first dataset includes localisation measurements of RSSI and AoA with the Cartesian co-ordinates of the target used from the Ultra-Wideband. 
The total data collected is just under 8 minutes for both scenarios, which is further split into 5 sessions each. 
The dataset consists of a time stamp, 8 RSSI values, followed by 8 AoA estimates, a location estimate using Bluetooth triangulation and Kalman filtering. The Ultra-Wideband location estimate is used as ground truth in this set. 

\subsubsection{Ultra-Wideband/Motion Capture}
The second dataset contains the UWB measurement data and location estimates produced by the motion capture system. 
This set also includes 5 sessions of data for each test scenario, of which each session is about 1.5 minutes long; total dataset is 15 minutes long. 
This dataset consists of a time stamp, 4 CIR values, 4 PSA values, 4 distances, UWB location estimates and the motion capture location estimates. 

\subsection{Evaluation Protocol \& Results}
Evaluation, using the proposed framework shown in Figure \ref{fig:overview}, was performed through a leave-one-session-out cross-validation scheme. 
In the first instance, the combined sets for both use-cases are used (with multiple sessions for both walking and trolley scenarios) for testing all sets separately whilst training is performed on all the remaining sets. 
This experiment is repeated in a similar fashion for both BLE and UWB calibration through ICON and the results are shown in Figure \ref{fig:results}. 
It can be seen that the BLE performance (using triangulation through Kalman filtering as discussed in Section \ref{kalman}) can be significantly improved through the use of the proposed framework. 
The mean error in localisation through ICON is reduced by approximately 70\%. 
Similarly, in the case of UWB, the mean error is reduced by about 50\%. Note, all the results presented here are statistically significant; evaluated through a two-sample Kolmogorov-Smirnov test. 

In the second experiment, we also evaluated the performance of the proposed framework across the two use-cases of \emph{walking} and \emph{trolley}. 
The Models were separately trained in a similar fashion as before, however focusing only on specific use-cases. This involved testing for individual sessions within a use-case and comparing against the baselines. 
Results for individual use-cases are summarised in Table \ref{tab:results}, where mean errors in localisation are reported.

As expected, walking scenario is relatively noisier and therefore there is generally a drop in errors in the case of trolley. 
However, this drop is exaggerated in the case of UWB where almost 50\% reduction in errors is observed. 
Using the proposed approach for UWB across two scenarios is still significantly more accurate, however less so in the case of Trolley.
For BLE, there is little difference between the two scenarios.

\begin{table}[tp]
\caption{Comparison of the two test scenarios (walking and trolley) against the corresponding baselines for both BLE and UWB calibration.
}
\centering
\resizebox{0.8\columnwidth}{!}{%
\begin{tabular}{lcc cc}
\toprule
 & \multicolumn{2}{c}{Walk} & \multicolumn{2}{c}{Trolley} \\
\cmidrule(lr){2-3} \cmidrule(lr){4-5}
& Baseline & ICON & Baseline & ICON \\
\midrule
BLE & $1.1083$m & $0.6670$m & $1.0530$m & $0.6576$m \\
UWB & $0.3238$m & $0.1184$m & $0.1465$m & $0.1038$m \\
\bottomrule
\end{tabular}}
\label{tab:results}
%\vspace{-0.5cm}
\end{table}

\section{Discussions}
In this paper, we have shown that calibration through ICON shows gains across multiple different technologies, and use-cases. 
Good results were observed despite a short pre-deployment campaign that was intentional in order to keep the calibration efforts minimal, and the off-line machine learning model training quick and efficient. 
For example, in the context of large retail shops, where, if such a framework is deployed, re-calibration would be quick to perform, unlike traditional finger printing.

We also observed, that the calibrating technology must have sufficiently greater accuracy than the underlying approach that is being calibrated.
This is very intuitive and thereby good news for commodity technologies like BLE due to its pervasiveness and penetration in large scale IoT deployments.
In the context of battery-powered UWB calibration, it is worth noting that they are easy to deploy in a plug-and-play fashion with minimal effort compared against the more accurate optical motion tracking system that requires a comprehensive physical setup prior to deployment. 
Improving UWB localisation through ICON can enable deployments in new applications which may have been previously unsuitable through traditional approaches. 

In this paper, we also used a single-layer neural network, mainly for two reasons: \emph{i)} efficient off-line training and, \emph{ii)} faster deployment.
It may be necessary in other contexts where deep neural networks will be required for modelling more dynamic environments for novel applications. Although, this may significantly increase the off-line training time.

\paragraph*{Limitations} In this paper, we have provided a benchmark study in the use of AI for calibrating various technologies under two specific use-cases. 
Real-world deployments may require modelling in the context of application-specific environments. 
It would be an interesting future subject to evaluate the performance of ICON in new environments but particularly in deployment scenarios that change over time. 
Such environments may require an extra effort during calibration or an automated re-calibration strategy for seamless model updates.

%Calibration shows gains across multiple different use-cases.. good results despite short pre-deployment campaign (small dataset). we intentionally did it for small datasets to minimize calibration effort for example for large retails parks and warehouses, which this research is motivated from... . not like full-fledged finger printing....

%if calibrating technology is more accurate, higher gains. very intuitive. yet a good news for commodity technologies like BLE because of their pervasiveness and penetration in smart phone. It is worth noting that battery-powered UWB (compared to optical systems) is easy to deploy on Ad Hoc basis.plug and play fashion. minimal effort.

%For this particular, shallow neural network were used. which are more efficient. for more realistic and dynamic environments, DNNs might be more suitable. 

%Limitation: we have not measured ... if environment change from calibration to actual deployment,  it would be interesting to evaluate, subject of our future work. 

\section{Conclusion}
Real-time localisation and tracking systems suffer from inaccuracies caused by the complex environments in which they are deployed.  This paper has presented a new approach to solving these issues using the ICON calibration framework based on an AI-driven approach. Through a generalisable localisation testbed, we demonstrated the performance of the ICON framework for BLE calibration through UWB and UWB calibration through a motion capture system across two use-cases.
We showed that the proposed approach is able to significantly reduce localisation errors by approximately 70\% and 50\% for BLE and UWB localisation, respectively.

%\section*{Acknowledgement}

\bibliographystyle{IEEEtran}
\bibliography{bare_conf}

\end{document}